\documentstyle[aps,preprint]{revtex}
\tighten

\title{Twisted sectors in three-dimensional gravity}

\author{M\'aximo Ba\~nados   }  
\address{Departamento de F\'{\i}sica Te\'orica, Universidad de
Zaragoza, \\ Ciudad Universitaria 50009, Zaragoza, Spain.}

\begin{document}

\maketitle

\begin{abstract}

Twisted sectors --solutions to the equations of motion with
non-trivial monodromies--  of three dimensional Euclidean gravity are
studied. We argue that upon quantization this new sector of the
theory provides the necessary (and no more) degrees of freedom to
account for the Bekenstein-Hawking entropy of three-dimensional black
holes. \\


\end{abstract}


\section{Introduction}

The asymptotic form of three-dimensional anti-de Sitter space,  
\begin{equation}
ds^2 \sim  r^2 dw d\bar w + {dr^2\over r^2}, \ \ \ \ \ \ \  
\label{aads0}
\end{equation}
enjoys remarkable properties. (Here $w$ is a complex coordinate
related to the spacetime real coordinates $t$ and $\varphi$ by, 
\begin{equation}
w=\varphi+ it.
\label{w}
\end{equation}
Note that we work in the Euclidean sector.) The most general set of
diffeomorphism leaving (\ref{aads0}) invariant form the conformal
group in two dimensions.  This symmetry was discovered by Brown and
Henneaux \cite{BH} more than 10 years ago, however, only recently it
was realized \cite{Strominger97} (see also \cite{Birmingham-}) that
it plays a central role in the statistical mechanical description of
three dimensional \cite{BTZ,BHTZ} black holes.

The metric (\ref{aads0}) is a solution to Einstein equations in three
dimensions with a negative cosmological constant $\Lambda = -2/l^2$.
The action for Euclidean three-dimensional gravity can be written  in
the convenient form,
\begin{eqnarray}
I_c[g] = {c \over 24\pi } \int \sqrt{g} (R + 2) d\tilde x^3,
\label{I}
\end{eqnarray}
where the constant $c$ is the Brown-Henneaux central charge \cite{BH},
\begin{equation}
c = {3l \over 2G},
\label{c}
\end{equation}
and the coordinates $\tilde x^\mu = x^\mu/l$ are dimensionless. In
this paper we shall refer all constants to $c$. 

The result discovered in \cite{BH} states that the asymptotic group of
symmetries of (\ref{aads0}) is the conformal group generated by two
copies of the Virasoro algebra\footnote{\label{L0'} This form of the
Virasoro algebra is appropriated to the anti-de Sitter sector with
$SL(2,C)$ as an exact symmetry.  In the black hole calculations we
shall encounter a shifted zero mode $L_0'=L_0-c/24$ appropriated to a
torus whose exact symmetries are $U(1)\times U(1)$ \cite{Coussaert-},
and the central term will have the $n^3\delta_{n+m}$ form.},
\begin{equation}
[L_n,L_m] = (n-m) L_{n+m} + {c \over 12} n(n^2-1) \delta_{n+m,0}
\label{Virasoro}
\end{equation}
where the central charge $c$ is the parameter appearing in the action
(\ref{I}).

The first attempt to relate the conformal properties of asymptotic
anti-de Sitter space with quantum black holes was done in
\cite{Carlip-T}. It was pointed out in that reference that the
degeneracy of states --treating $L_0$ as a number operator-- was
proportional to the black hole area but with a different coefficient.
Strominger \cite{Strominger97} made the striking observation
that if one regards the conformal algebra as being generated by a 1+1
unitary conformal field theory, as suggested by the adS/CFT
correspondence \cite{Maldacena,Gubser-,Witten98}, then the degeneracy
of states for large values of $L_0$ and $\bar L_0$ is exactly equal to
the Bekenstein-Hawking entropy for the corresponding black hole with
mass $Ml = L_0+\bar L_0 - c/12$ and spin $J=L_0-\bar L_0$. (See
\cite{Carlip98-2} and \cite{Solodukhin} for recent extensions of this
idea to higher dimensions.)

In a purely gravitational calculation (like the one suggested in
\cite{Carlip-T}), the boundary CFT is Liouville theory \cite{CHvD}
which is a single field and has an effective central charge equal to
one. This means that the degeneracy of states grows as
\cite{Kutasov-,Carlip98} exp$(2\pi\sqrt{L_0/6})$ instead of the
desired Cardy form exp$(2\pi\sqrt{c L_0/6})$ which would give the
right result. This happens because, contrary to a system of D free
bosons
with central charge $D$ and a degeneracy growing with $D$, in
Liouville theory the central charge is related to the coupling, not
the number of fields. Accordingly, the degeneracy does not grow with
$c$. In order to have the right counting one would need to have ``$c$
Liouville fields". This has lead many authors to conclude that the
gravitational degrees of freedom carry only the thermodynamical
aspects of the Bekenstein-Hawking entropy \cite{Martinec,Hyun,Giveon-}
(see \cite{Jacobson} for an earlier discussion of this idea). 

In this paper we will show that if one includes solutions to the
equations of motions with twisted boundary conditions (on the torus)
the degeneracy grows and provides the necessary degrees of freedom to
account for the Bekenstein-Hawking entropy. We will show that
the degeneracy in the twisted sector is, 
\begin{equation}
\rho_q(L_0)=\exp\left(2\pi\sqrt{q L_0/ 6}\right)
\label{deg0}
\end{equation}
where $q$, the order of the twisting, is a positive integer. On the
other hand, unitarity in the twisted sector will restrict $q$ by $c/q
\geq 1$. Thus, the sector with the maximum degeneracy $q=c$ gives the
right Bekenstein-Hawking value. (Note that the sum over $q$ is
dominated by its maximum value $q=c$.)

It is important to stress here that the formula (\ref{deg0}) has
nothing to do with the Cardy formula (which requires modular
invariance and unitarity). Eq. (\ref{deg0}) will follow by counting
states explicitely in a given representation of the Virasoro algebra.
(In our situation, the sum over representations give
subleading contributions to the degeneracy.) 

The results of this paper provide a geometrical justification for the
results reported in \cite{B4}.

\section{Classical solutions in adS$_3$ gravity}
\label{2}

In this section we describe the structure of the classical space of
solutions (untwisted and twisted) of three-dimensional anti-de Sitter
gravity.   

\subsection{AdS$_3$ and Riemann surfaces}  
\label{Riemann}

The metric (\ref{aads0}) with $w=\varphi+it$ is defined on the
``cylinder" because $\varphi$ is an angle. It will be useful for our
purposes to work on the plane. This is achieved by performing a
conformal transformation which also serves as an example to illustrate
the conformal properties of (\ref{aads0}). Consider the transformation
from the cylinder $w$ to the plane $z$ via,
\begin{eqnarray}
z &=& e^{-iw} \left( 1 +  {1\over r^2}+ ... \right)\nonumber\\
r' &=& e^{i(w-\bar w)/2} \, r \label{exp}
\end{eqnarray}
It is direct to see that the new metric in the coordinates $\{z,r'\}$
looks, to leading order in $r$, exactly like (\ref{aads0}).
This is an example of the invariance of (\ref{aads0}) under a
{\it finite} conformal transformation. Note that the coordinate $r$ is
not invariant under the transformation. In fact, from (\ref{exp}) we
see that $r^2$ transforms like a primary with conformal dimension
$(1,1)$.  Of course this is necessary if one expects $r^2 dw d\bar w$
to be invariant.    

It will be convenient to introduce a new radial coordinate $\rho$
defined as $r' = e^\rho$. This change, plus the conformal
transformation (\ref{exp}) bring (\ref{aads0}) to the form, 
\begin{equation}
ds^2 \sim e^{2\rho} dz d\bar z + d\rho^2.
\label{aads}
\end{equation}
This asymptotic metric will be our starting point.

The metrics (\ref{aads0}) and (\ref{aads}) are actually exact
solutions to the field equations, although their global properties
differ. The metric (\ref{aads0}) corresponds to the vacuum black hole
\cite{BTZ,BHTZ}, while (\ref{aads}) to anti-de Sitter space. This can
be seen by setting $\lambda = 1/r$ in (\ref{aads0}) and $\lambda=
e^{-\rho}$ in (\ref{aads}). Both metrics are mapped into the upper
Poincare plane. Since $\varphi$ in (\ref{aads0}) is compact there are
identifications in this case. The corresponding solution is the vacuum
black hole. On the other hand, the coordinate $z$ in (\ref{aads})
lives in the whole complex plane $z=x+iy$, and the corresponding
solution is anti-de Sitter space without any identifications. This
change in the topology occurs because in the transition from
(\ref{aads0}) to (\ref{aads}) via (\ref{exp}) we have kept only the
leading terms, ignoring Schwarztian derivative pieces which precisely
relate anti-de Sitter space with the vacuum black hole through the
shift $L_0 \rightarrow L_0 - c/24$. 

The three dimensional black hole is asymptotically anti-de Sitter and
it approaches (\ref{aads}) at infinity.  In the Euclidean sector, the
time coordinate is periodic in order to avoid conical singularities,
$t\sim t + \beta$. The topology in the $w$ plane is a torus with the
identifications $w\sim w+ 2\pi n +2\pi m\tau$ ($n,m\in Z$) where
$\tau$ is related to the black hole temperature and angular velocity
\cite{BBO,Maldacena-}. In the $z$-plane this identification reads, 
\begin{equation}
z \sim e^{-2\pi i \tau} z.
\label{zz}
\end{equation}
The identification (\ref{zz}) can also be introduced in the anti-de
Sitter sector and leads to thermal anti-de Sitter space (TadS)
\cite{Hawking-Page}. In three dimensions, thermal anti-de Sitter space
is related to the black hole via a modular transformation
\cite{Maldacena-}. 

The Euclidean sector of three-dimensional anti-de Sitter space is
particularly interesting because the mathematics of Riemann surfaces
is available. One can consider generalizations of (\ref{aads}) on
which the coordinate $z$ is defined on a Riemann surface with genus
$g$, and the conformal group will still act in a natural way. The
solutions to the equations of motion can be summarized in the table,
\begin{equation}
\begin{array}{c|c} 
2d-\mbox{surface}            & 3d-\mbox{solution} \\
\hline 
 \mbox{sphere }(g=0)         & \mbox{adS}_3  \\
\mbox{torus }(g=1)           & \mbox{black holes / TadS} \\
\mbox{Riemann surface }(g>1) & \mbox{?}
\end{array}
\label{table0}
\end{equation}
(Note that in each case there may be conical or cusp singularities, if
sources are added.)  We shall see that admitting solutions with
non-trivial monodromy properties leads to higher genus solutions, but
we shall not discuss this point in any detail.  To our knowledge
solutions to the Euclidean equations of motion with higher genus (a
solid Riemann surface) have not appeared in the literature. (See
\cite{Aminneborg2-} for a recent discussion about solutions with
higher genus in Minkowskian signature.)

\subsection{The untwisted solution}  

A key property of three-dimensional gravity is the absence of bulk
degrees of freedom. This makes possible to find the general solution
to the equations of motion. Under the boundary conditions
(\ref{aads}), the general solution  has the form of a ``travelling
wave", 
\begin{equation}
 ds^2 = f  dz^2 + \bar f   d\bar z^2 +
\left(e^{2\rho} + f\bar f e^{-2\rho}  \right)dz d\bar
z + d\rho^2,
\label{dsTT}
\end{equation}
where $f=f(z)$ and $\bar f=\bar f(\bar z)$ are arbitrary functions of
their arguments \cite{Baires}. Locally, this metric has constant
curvature (as demanded by the vacuum three-dimensional Einstein
equations) and, for $\rho \rightarrow \infty$, it approaches
(\ref{aads}) with the correct fallof behaviour \cite{BH}.   

To prove that (\ref{dsTT}) is the most general solution we resort to
the analysis of \cite{BH}. The most general perturbation of the
background metric (\ref{aads}) which preserves its asymptotic
structure is constructed by acting on it with a set of asymptotic
conformal vectors $\xi^\mu= \xi^\mu(\varepsilon,\bar \varepsilon)$
which depend on two arbitrary functions $\varepsilon(z)$ and $\bar
\varepsilon(\bar z)$. Acting with these vectors on (\ref{aads}) one
produces the solution (\ref{dsTT}) with,
\begin{equation}
f(z) = {6 \over c}\, T(z),
\label{fT}
\end{equation}
where $c$ is given in (\ref{c}). The same formula holds for the
anti-holomorphic side. We have named $f(z)$ as in (\ref{fT}) because
the function $T(z)$ transform under the above symmetry as a Virasoro
operator. In the following we shall denote the metric (\ref{dsTT}) as
$ds^2(T,\bar T)$.  

Actually, this construction yield only the pieces linear in $f$ and
$\bar f$ of (\ref{dsTT}) \cite{Navarro-}. To obtain the full exact
solution (\ref{dsTT}) we simply note that the term $f\bar f
e^{-2\rho}dz d\bar z$ --which makes (\ref{dsTT}) an exact solution--
does not affect the leading and first subleading terms. Since
three-dimensional gravity does not have any bulk degrees of freedom,
it follows that (\ref{dsTT}) is the most general solution with the
boundary condition (\ref{aads}). 

The construction of (\ref{dsTT}) by acting on (\ref{aads}) with the
conformal generators parallels the way one constructs representations
of the Virasoro algebra. The analogue of the vacuum state $|0\rangle$
is the anti-de Sitter background (\ref{aads}) which is $SL(2,C)$
invariant. Acting on it with the conformal generators one finds the
general solution (\ref{dsTT}).  The metric (\ref{dsTT}) is then a
``descendant" of the metric (\ref{aads}) which is the ``primary". We
shall make this point precise below in this section and in Sec.
\ref{Quantization}.  

Generically, the perturbations produced on the anti-de Sitter metric
(\ref{aads}) via conformal transformations are defined only locally
because, apart from the $SL(2,C)$ transformations for which
(\ref{aads}) is invariant, all other conformal mappings are not
globally well-defined. In this sense, the metric (\ref{aads}) is a
background solution whose local perturbations, consistent with the
equations of motion, are described by (\ref{dsTT}). In other words,
since (\ref{dsTT}) has constant curvature, it can be obtained from
(\ref{aads}) via identifications with some discrete group. However,
this discrete subgroup will act freely only on some special cases. 

The ``vacuum" metric with $T=\bar T=0$ is not the only possible
background solution. There exists a continuum of metrics for which
$T(z)$ has the particular form,  
\begin{equation}
T(z) = {\triangle \over z^2},
\label{global}
\end{equation}
where $\triangle$ is some constant. These solutions are also globally
well-defined and for suitable values of $\triangle$ they represent
black holes. See \cite{Baires} for the transition from a metric with
$T(z)$ of the form (\ref{global}) to a 3d black hole of mass
$M=\triangle+\bar \triangle-c/12$ and spin $J=\triangle-\bar
\triangle$. We shall denote this particular set of solutions as
$ds^2(\triangle,\bar \triangle)$.  

Before leaving this section, we note that the asymptotic conformal
symmetries of (\ref{aads}) can be ``lifted" to act on the space of
solutions (\ref{dsTT}). Let $g_{\mu\nu}(T,\bar T)$ the metric
(\ref{dsTT}) and $x^\mu=\{z,\bar z, \rho\}$. There exists vector
fields $\xi^\mu_{\epsilon,\bar\epsilon}(x)$ depending on two arbitrary
(anti-) holomorphic functions $\epsilon(z)$ $(\bar\epsilon(\bar z))$
such that under the change $x^\mu \rightarrow x'^\mu = x^\mu +
\xi^\mu(x)$ the metric (\ref{dsTT}) transform as, 
\begin{equation}
g_{\mu\nu}(T,\bar T) + \delta g_{\mu\nu} = 
g_{\mu\nu}(T+\delta T,\bar T + \delta \bar T) 
\label{deltag}
\end{equation}
with  
\begin{eqnarray}
\delta T = \epsilon \partial T + 2 \partial \epsilon T + {c
\over 12} \partial^3 \epsilon,
\label{deltaT}
\end{eqnarray}
and $c$ is given in (\ref{c}). See \cite{Baires} for the explicit form
of the residual conformal vectors which, of course, coincide
asymptotically with the vectors found in \cite{BH}.  This symmetry is
an infinite dimensional symmetry of the space of solutions. In the
language of representations of the Virasoro algebra, it maps the
members of a conformal family into themselves.  

Suppose we start with the metric (\ref{aads}) and perform
a finite conformal transformation $z \rightarrow z'(z)$. The finite
form of (\ref{deltaT}) is known and involves a Schwarztian derivative.
Since (\ref{aads}) can be regarded as a particular case of
(\ref{dsTT}) for which $T(z)=0$, making this finite transformation
will give a solution of the form (\ref{dsTT}) with $T'(z') = (c/12)
\{z,z'\}$, where $\{,\}$ denotes the Schwarztian derivative. Defining 
$A(z')=\ln (\partial z/\partial z')$ we find the
Liouville type stress tensor,
\begin{equation}
T'(z') = {c \over 24}(- (\partial A)^2 + 2 \partial^2 A). 
\label{A'}
\end{equation}
The function $A$ can be identified with the holomorphic function
appearing in the general solution to the Liouville equation.  See
\cite{Navarro-} for a related approach to obtain the Liouville stress
tensor. 

Since the residual symmetries of (\ref{dsTT}) are nothing but changes
of coordinates they can be generated by the constraints of general
relativity, supplemented with appropriated boundary terms. In the
gauge fixed form of the space of solutions (\ref{dsTT}), these
generators are the functions $T$ and $\bar T$ and the associated
algebra is the Virasoro algebra \cite{BH}. In terms of the modes $L_n$
defined in the usual way,
\begin{equation}
T(z) = \sum_{n\in Z} {L_n \over z^{n+2} },
\label{Ln}
\end{equation}
one obtains the algebra (\ref{Virasoro}).

An important property of these transformations is that $e^\rho$ is not
a scalar. Indeed  $e^{2\rho}$ is to leading order a primary field
with conformal dimension $(1,1)$ and hence $e^{2\rho} dz d\bar z$ is
invariant. This also explains the presence of the term $e^{-2\rho} T
\bar T$ in the solution (\ref{dsTT}).  Since $T$ and $\bar T$ are
quasi-primaries with conformal dimensions $(2,0)$ and $(0,2)$
respectively, the product $T\bar T$ has dimension $(2,2)$. The
combination $e^{-2\rho} T \bar T$ then has dimension $(1,1)$, as
needed.

\subsection{The twisted sector} 

Motivated by the problem of black hole entropy --lack of enough states
to account for the large black hole degeneracy, see
\cite{Carlip98,Martinec,Baires} for recent discussions-- our goal in
this section is to generalize the space of solutions described in the
last section.  

The generalization we have in mind is to admit non-trivial monodromies
in the angular direction. In other words, we consider solutions of the
form (\ref{dsTT}) for which the function $f(z)$ has the property, 
\begin{equation}
f(e^{2\pi i}z) = g \cdot f(z),
\label{g}
\end{equation}
where $g$ is a symmetry of the space of solutions. In Sec.
\ref{Quantization} we shall make this notation precise. For the time
being, we consider functions $f(z)$ which have fractional powers in
the variable $z$. On
the torus (black hole) the angular cycle is non-contractible and thus
the twisting does not introduce singularities. On the sphere (adS)
there will be two singularities in the north and south poles of the
Riemann sphere.  

We shall argue in Sec. \ref{CS} that in the Chern-Simons formulation
of three-dimensional gravity the twisted solutions arise in a
completely natural way. 

We consider the solution (\ref{dsTT}) with a mode expansion for
$f(z)$ having fractional powers of $z$.  Let $q$ be a non-negative
integer, which will be called the order of the twisting.  We replace
(\ref{fT}) with
\begin{equation}
f(z) = {6 \over c}\, q T(z)
\label{fTq}
\end{equation}
where  
\begin{eqnarray}
T(z) = {1 \over q} \sum_{r=0}^{q-1} \sum_{m\in Z} { L_{m+r/q} \over
          z^{n+r/q+2}}.   
\label{Lnr}
\end{eqnarray}
The factors $q$ and $1/q$ appearing in (\ref{fTq}) and (\ref{Lnr})
are conventional and included for future convenience. Similar
formulae are assumed for the anti-holomorphic factor. The stress
tensor $T(z)$ is  a finite sum of the form, 
\begin{eqnarray}
qT(z) = T_0(z) + T_1(z) + \cdots + T_{q-1}(z), 
\end{eqnarray}
where $T_r(e^{2\pi i}z)=e^{2\pi i r/q}T_r(z)$. Clearly
$T_0(z)$ corresponds to the untwisted solution mentioned in the
previous paragraph. The parameter $q$ is regarded as a new degree of
freedom in the theory and we shall sum over it. 

The modes $L_{n+r/q}$ can be inverted in terms of $T(z)$ in two steps. 
First we note that $T_r$ can be expressed in terms of $T$ as,
\begin{equation} 
T_{r}(z) =  \sum_{s=0}^{q-1} T(e^{2\pi s}) e^{2\pi i sr/q},
\label{Tr}
\end{equation}
and the modes $L_{n+r/q}$ are given in terms of $T_r$ as,
\begin{equation}
L_{n+r/q} = \oint {dz \over 2\pi i} z^{n+r/q+1} T_{r}(z).
\label{modesr}
\end{equation}

The metric (\ref{dsTT}) is still an exact solution because the field
equations only see the local properties. For the same reason, the
modes $L_{m+r/q}$ satisfy the same Virasoro algebra (\ref{Virasoro}) 
but now the label $n$ in (\ref{Virasoro}) takes values on $Z+ r/q$.
One finds the twisted Virasoro algebra, 
\begin{eqnarray}
[L_{n+r/q},L_{m+s/q}] &=& (n - m + (r-s)/q) L_{n+m +(r+s)/q}
\nonumber\\ && + {c \over 12} (n+r/q)[(n+r/q)^2-1]
\delta_{n+m+(r+s)/q}.
\label{orb}
\end{eqnarray}
This algebra has appeared in \cite{Borisov-} in the context of cyclic
orbifolds \cite{Klemm-}.

Note that the modes $L_{n+r/q}$  are defined such that $0\leq r<q$. 
The algebra (\ref{orb}) should be understood with the periodicity
and reality conditions
\begin{eqnarray}
L_{n+ (r+q)/q} &=& L_{n+1 + r/q}, \\
L_{n - r/q} &=& L_{n-1 + (q-r)/q}, \\ 
L_{n+r/q}^\dagger &=& L_{-n-r/q}.
\label{perio}
\end{eqnarray} 
The algebra (\ref{orb}) has a Virasoro sub-algebra (for $r=0$)
generating the single-valued conformal transformations of $z$ with
central charge $c$. It is clear that this sub-algebra is isomorphic to
the Brown-Henneaux generators $L_n$ acting on the single valued, or
untwisted, sector. 

There is an important point here concerning the value of the central
charge appearing in (\ref{orb}).  If one computes the algebra, or OPE
of the function $T(z)$ defined in (\ref{Lnr}) using (\ref{orb}), one
finds a central charge $c/q$. This is correct. The function $T(z)$
contains all sectors $T_r(z)$, $r=0,1,.., q-1$. It is the trivial
monodromy generator $T_0(z)$ what needs to be matched with the
Brown-Henneaux one, not the full $T(z)$. $T_0(z)$ defines a subalgebra
of the full $T(z)$ and satisfy the Virasoro algebra with central
charge $c$, as desired. This point will have an important consequence
below.     

The perturbed metric (\ref{dsTT}), for a general $T(z)$ with twisted
contributions, is no longer periodic under $z\rightarrow e^{2\pi i}z$.
For the purposes of this paper, this classical non-periodicity will be
of no importance because, as we discussed above, the only globally
defined metrics are those for which $T(z)$ is of the form
(\ref{global}), for some value
of $\triangle$. These metrics are clearly periodic. The constant
$\triangle$ (equal to the conformal dimension of the state) is made up
of periodic and non-periodic perturbations, but the observable
globally defined metric is periodic. We shall come back to this point
in the next section.

\section{Quantization.}
\label{Quantization}

The goal of this section is to study the quantum version of the space
of solutions described in the last section.  

\subsection{The spectrum. } 
\label{Spectrum}

The space of solution described by metrics of the form (\ref{dsTT}) is
infinite dimensional, and so far, completely disorganized. Any
function $f(z)$, say $f(z) = z^3$ or $f(z)=z^{1/2}$, provides a
solution and there is no way to predetermine their physical
properties. 

Since there exists a Virasoro symmetry acting on (\ref{dsTT}) mapping
solutions into solutions, one can classify the solutions using the
representations of the Virasoro algebra. In particular, this will
enable us to select the physically reasonable ones, e.g., with
positive energy. 

The idea is to find a map from the vectors in the representation space
of the Virasoro algebra, and the space of solutions (\ref{dsTT})
\cite{Baires}. We start by identifying the metric associated to the
vacuum state $|0\rangle$. There is a unique candidate to represent
this state, namely the anti-de Sitter metric (\ref{aads}).

As we mentioned in Sec. \ref{Riemann}, the metric (\ref{aads}) is an
exact solution to the field equations. In fact, it represents
Euclidean anti-de Sitter space without any identifications. The metric
(\ref{aads}) is invariant under global $SL(2,C)$ transformations and
it is a particular case of (\ref{dsTT}) on which $T$ and $\bar T$ are
equal to zero. We are then led to the identification,
\begin{equation}
|0\rangle \ \ \ \ \leftrightarrow \ \ \ \
ds^2(0,0)= e^{2\rho}dzd\bar z + d\rho^2.
\label{ads-0}
\end{equation}

The next step is to identify the ``primaries" from which all other 
representations will be constructed. 

Other globally well-defined solutions are those for which $T(z)$
is of the form (\ref{global}). Consider the metric (\ref{dsTT}) with
$f(z)$ given in (\ref{fTq}) and $qT=h/z^2$ where $h$ is a positive
real number (the same for the anti-holomorphic side). We denote this
metric as $ds^2(h,\bar h)$. We identify the primary fields $|h,\bar
h\rangle$ with the background solutions of the form (\ref{global}), 
\begin{equation}
|h,\bar h\rangle \ \ \ \ \leftrightarrow  \ \ \ \    ds^2(h,\bar h).
\label{dshh}
\end{equation}

The metrics $ds^2(h,\bar h)$ are isometric to black holes or conical
singularities \cite{Deser}, depending on the values of $h$ and $\bar
h$ \cite{Baires}.  In other words, the three-dimensional topologies
associated to $ds^2(h,\bar h)$  depend on the values of $h$ and $\bar
h$. In complete analogy with Liouville theory, the conformal
dimensions of the primaries $h$ and $\bar h$ are separated in various
sectors (see also \cite{Martinec,Frolov-FGK,Navarro-N2}): 
\begin{equation}
\begin{array}{lcl} 
h=\bar h=0       &  &\mbox{anti-de Sitter space} \\   
0<h+\bar h <c/12, \ \ \ &  &\mbox{conical singularities} \\
h=\bar h= c/24    &  &\mbox{vacuum black hole}  \\
h+\bar h>c/12,\ \ \ \ & & \mbox{black holes}  
             \end{array}
\label{spectrum}
\end{equation}
Having defined the primaries, a full representation is constructed by
acting with the conformal generators. 

The above construction can be summarized and completed by the single
formula
\begin{equation}
T_\Psi(z) = \langle\Psi| \hat T(z) |\Psi\rangle,
\label{map}
\end{equation}
where $|\Psi\rangle$ is any normalized state in a given representation
of the Virasoro algebra. 

We identify the function $T(z)$ appearing in (\ref{dsTT}) with
the expectation value of the quantized operator $\hat T$ on a given
state $|\Psi\rangle$; $T(z)=T_\Psi(z)$. In other words, instead of
considering (\ref{dsTT}) with arbitrary values for $T(z)$, we consider
only those functions $T_\Psi(z)$ which are the expectation values of
the quantised $\hat T(z)$ on some state $|\Psi\rangle$. This still
leaves an infinite dimensional space of solutions, but they
are now nicely organized according the the representations of the
Virasoro algebra. In particular, if we only consider unitary
representations, positivity of the ADM mass is guaranteed.

Note that since conformal transformations acting on the state
$|\Psi\rangle$ produce conformal transformations on $T_\Psi(z)$ of the
form (\ref{deltaT}), the identification (\ref{map}) implements
correctly the mapping of solutions into solutions via (\ref{deltaT}).
Indeed, consider an infinitesimal conformal transformation
$z\rightarrow z + \epsilon(z)$.  This transformation has an
associated group element $U_\epsilon = 1 + i \oint \epsilon(z) T(z)$
acting on the representation space: $|\Psi\rangle \rightarrow
|\Psi'\rangle = U_\epsilon |\Psi\rangle$.  It is direct to see that
the effect of this transformation on $T_\Psi$ is exactly equivalent to
(\ref{deltaT}). 

The map (\ref{map}) can be extended to the full metric. We extend the
above analysis to the anti-de holomorphic sector. To
each state in the full Fock space $|\Psi\rangle$ there is a
corresponding solution to the equations of motion \cite{Baires},
\begin{equation}
ds^2_\Psi = \langle\Psi| d\hat s^2 |\Psi\rangle,
\label{gmap}
\end{equation}
where $d\hat s^2$ is the metric (\ref{dsTT}) on which $T$ and $\bar T$
have been replaced by their quantized versions $\hat T$ and $\hat{\bar
T}$.  Note that the metric (\ref{dsTT}) does not involve products of
non-commuting operators \footnote{This property will not survive in a
full gauge invariant quantization. Note, however, that in a
Chern-Simons formulation one would need to regulate at most a
quadratic function of the currents (Sugawara type).} and thus is
well-defined as an operator. 

The statistical mechanical picture which follows from the above
analysis is the following.  We construct the representations of the
twisted Virasoro algebra (\ref{orb}). Then we seek (and count)
those states which under the map (\ref{gmap}) yield a black hole with
a given mass and spin.  We shall show that the number of these states
is enough to account for the Bekenstein-Hawking degeneracy.  

Before going to the representations of the algebra (\ref{orb}), let us
study the effect of the twisting on the quantum metric. From the
algebra (\ref{orb}) it is direct to prove the identity, 
\begin{equation}
e^{-\alpha (n+r/q)} L_{n+r/q} = e^{\alpha L_0} L_{n+r/q}\, e^{-\alpha
L_0},  
\label{id}
\end{equation}
where $\alpha$ is a complex number\footnote{The proof is as follows.
We consider $X(\alpha) = e^{\alpha L_0} L_{n+r/q}\, e^{-\alpha
L_0}$. Using the algebra (\ref{orb}) it is direct to see that $X$
satisfies the differential equation $dX/d\alpha = -(n+r/q)X$ whose
solution is $X = e^{-\alpha(n+r/q)} X_0$. The constant $X_0$ does not
depend on $\alpha$ and thus can be evaluated on $\alpha=0$ yielding
$X_0 = L_{n+r/q}$.}. Using this formula with $\alpha=2\pi i$ one can
check that $\hat T(z)$ given in (\ref{Lnr}) transforms as (here
$T(z)=T(z)dz^2$ is understood as a quadratic differential),  
\begin{eqnarray}
\hat T(e^{2\pi i} z) &=& A \, \hat T(z) \, A^{-1},
\label{A} 
\end{eqnarray}
where $A = e^{2\pi i L_0}$ is the generator of the cycle,
\begin{equation}
z\rightarrow e^{2\pi i}z,
\label{2pi}
\end{equation}
acting on Fock space. In this sense the metric (\ref{dsTT}) as an
operator is well-defined.  

The classical function $T_\Psi(z)$ appearing in the classical solution
(\ref{dsTT}) transforms as,
\begin{equation}
T_\Psi(e^{2\pi i}z) = T_{\Psi'}(z),
\label{2pi'}
\end{equation}
where $\Psi'$ is the transported state, 
\begin{equation}
|\Psi\rangle \rightarrow |\Psi'\rangle= e^{2\pi i L_0}|\Psi\rangle.
\label{psi'}
\end{equation}
The particular set of states for which $T_\Psi$ is invariant under
(\ref{psi'}), $T_\Psi = T_{\Psi'}$, yield a well-defined univalued 
metric. These states are those for which $L_0$ is diagonal because the
transportation (\ref{psi'}) gives a phase not seen by the expectation
value $T_\Psi(z)=\langle\Psi| T(z) |\Psi \rangle$.  As expected, the
classical solutions associated to these states are the globally
well-defined metrics with $T(z)$ given in (\ref{global}).

\subsection{Black hole states}
\label{Counting}

The space of states which diagonalize $L_0$ are those in the Verma
module, or linear combinations of them at a given level. These states
have the extra property that the associated classical metric,
through (\ref{gmap}), is univalued and isometric to
a black hole \cite{Baires} with a mass and spin given by 
\begin{equation}
M = L_0' + \bar L_0' \ \ \ \  J = L_0' - \bar L_0',
\label{MJ}
\end{equation}
where $L_0'=L_0-c/24$ (see footnote \ref{L0'}).  

We shall now count the number of states compatible with a given value
of $L_0'$ in a representation created by a ``primary" $|h\rangle$ of
the twisted algebra (\ref{orb}). Since we are interested in the
microscopic structure of black holes, we take as primary state the
vacuum black hole with $h=c/24$ and $L_0'=0$. There is a subtle issue
here
because the vacuum black hole does not have the same topology as the
massive black hole. We shall imagine that the primary is arbitrarily
close to be a vacuum black hole, but its topology will be that of a
massive black hole: a solid torus. In particular, the boundary
described by the coordinate $z$ is a torus. 

For $h=c/24$ and $c>1$ (and $q\leq c$, see below), the representations
of (\ref{orb}) are unitary and irreducible. The excited levels are
given by the states    
\begin{eqnarray}
|n_i,r_i\rangle &=& L_{-n_1 + r_1/q} \cdots L_{-n_s+r_s/q}|c/24\rangle 
\end{eqnarray}
where $n_i=1,2,3,...$ and $r_i = 0,1,2,...,q-1$. These states satisfy
(assuming they have been normalized),
\begin{equation}
\langle n_i,r_i|L_{n+q/r}|n_i,r_i \rangle = \delta^0_n \delta^0_r L_0.
\end{equation}
The associated metric, through (\ref{gmap}), is then univalued and
globally defined with $T(z)$ of the form (\ref{global}) 
and $\triangle=L_0$.  

From (\ref{orb}), we deduce that $L_0$ is a number operator and its
value on the state $|n_i,r_i\rangle$ is given by,   
\begin{equation}
L_0' := L_0 - {c \over 24} = \sum_{i=1}^s(n_i-r_i/q).
\label{l0}
\end{equation}
Note that since we take the vacuum black hole as background metric 
with $h=c/24$, there is a direct relation between the excited levels
and $L_0'$ which is the quantity related directly to the black hole
mass and spin, as in (\ref{MJ}), with no added constants.  

The sum over $n_i$ in (\ref{l0}) is the contribution from
the untwisted sector which we know does not give the right degeneracy.
The second piece $r_i/q$ provides a `hyperfine structure' which
increases the degeneracy. 

Eq. (\ref{l0}) can be rewritten in the form, 
\begin{equation}
qL_0' = \sum_i m_i, \ \ \ \ \ \mbox{with} \ \ \ \ \ \ m_i = 
qn_i-r_i.
\label{qL0}
\end{equation}
The number of states compatible with a given value of $L_0$ is
given by the number of combinations of the pair $n_i,r_i$ such that
(\ref{qL0}) is satisfied. 

In what follows, the following (trivial) lemma will be of great help.
Let $n \in Z$ and $0\leq r <q$: 
\begin{eqnarray}
\mbox{{\it  The map:}  } \{n,r\} \rightarrow m = qn - r \nonumber
\\ 
\mbox{{\it is one to one (and invertible).}   } 
\label{lemma}
\end{eqnarray}
Given any $m\in Z$ there exists a unique pair $\{n, r\}$ such that $m
= qn-r$. Then, it follows that all posible
combinations of $\{n_i, r_i\}$ give all possible positive integers
$m_i$ once. The problem of how many states --varying $n_i$ and $r_i$--
can exist for a given $L_0'$ can then be formulated as the number of
ways that the integer $q L_0'$ can be written as a sum of integers
$m_i$. This is of course given by the Ramanujan formula,
\begin{equation}
\rho \sim e^{2\pi \sqrt{q L_0'/6}},
\label{Rama}
\end{equation}
which is valid for large values of $qL_0'$. Note the factor of $q$
multiplying $L_0'$. Since $q$ is an integer greater than one, the
degeneracy has certainly grown with respect to the pure Liouville
calculation which has an affective central charge equal to one. 
The key step in this counting is that even though $L_0'$ is not an
integer, the combination $qL_0'$ is.  
 
If we choose $q=c$, then (\ref{Rama}) gives the expected value for the
black hole entropy \cite{Strominger97}. Thus, in principle, we have
enough states but this is certainly not satisfactory. We would like to
prove that $q$ must not exceed $c$.  We shall now argue that for
integer values of the Brown-Henneaux central charge $c$, unitarity in
the representations of the algebra (\ref{orb}) restricts the possible
twistings and $q$ must indeed lie in the range,
\begin{equation}
1\leq  q\leq c. 
\label{qc}
\end{equation}
Thus, $q=c$ is the sector with the maximum degeneracy (maximum
twisting) and yield the correct Bekenstein-Hawking value. (Note that
the sum over $q$ is dominated by the highest twisting sector $q=c$.)

This condition arises from a re-ordering of the generators $L_{n+r/q}
\rightarrow Q_m$ ($m\in Z$) given by,
\begin{equation}
L_{n+r/q} = {1 \over q}Q_{qn+r} .
\label{LtQ}
\end{equation}
This is a purely algebraic step. We choose to enumerate the $L$'s by
integers subscripts according to (\ref{LtQ}). A geometrical
interpretation for the $Q$'s (on the sphere) will be given in next
section. 

In view of (\ref{lemma}), it is evident that the map (\ref{LtQ}) is
one to one. The set of all $Q_n$ with $n\in Z$ contains all
$L_{n+r/q}$ and viceversa.  Some examples of the above transformation
are: $Q_0 = qL_0$, $Q_1=qL_{1/q}$, $Q_q=qL_1$, etc.  The
transformation (\ref{LtQ}) is a particular case of a general procedure
studied in \cite{Borisov-}.  

The algebra of the modes $Q_n$ can be computed directly from the
algebra (\ref{orb}) and yields, 
\begin{equation}
[Q_n,Q_m] = (n-m) Q_{n+m} + {c/q \over 12} n(n^2-q^2) \delta_{n+m}.
\label{QQ}
\end{equation}
This algebra, up to a shift in the zero mode $Q_0$ that we discuss
below, is a Virasoro algebra with central charge $c/q$. 

It should now be clear where does condition (\ref{qc}) comes from.
Unitary representations for (\ref{QQ}) exist only for $c/q \geq 1$
thus leading to (\ref{qc}). 

Unitarity is also achieved for a central charge $c/q$ smaller than one
provided it has the Kac form $c/q=1-6/((m+2)(m+3))$ with $m=0,1,2,..$.
This condition yields a strange quantization condition for $c$, and
thus for Newton's constant. The degeneracy in this case is more
complicated to determined because there are null states in the Verma
module. Most likely, the counting will still give the correct
degeneracy provided one includes all conformal dimensions producing a
modular invariant partition function and using the Cardy formula.
Since the central charge is $c/q$ and $Q_0=qL_0$, one would have
$\sqrt{(c/q)Q_0}=\sqrt{cL_0}$, as desired. We shall not consider this
situation here. 

The form of the central term in (\ref{QQ}) may seem strange. Note that
the transformation (\ref{LtQ}) maps $L_0,L_{\pm 1}$  into
$Q_0,Q_{\pm q}$ generating an $SL^{(q)}(2,\Re)$ sub-algebra. Let us
shift
$Q_0$ such that the central term in (\ref{QQ}) acquires the usual form
with $Q_0,Q_{\pm 1}$ generating an $SL(2,\Re)$ algebra. This is
achieved by redefining the new $Q_0^{new}$, 
\begin{equation}
Q^{new}_0  = Q_0 - {c \over 24} \left(q - {1 \over q}\right). 
\label{shift}
\end{equation}
Let us take the sector $q=c$. In terms of this new zero mode, unitary
representations for (\ref{QQ}) exists provided $Q_0^{new}\geq 0$. From
(\ref{shift}) and (\ref{LtQ}) we obtain the bound for $L_0$
\begin{equation}
L_0 \geq {c \over 24} - {1 \over 24c}.
\end{equation}
 For $c$ large the first term dominates and this is the
condition that eliminates (almost all) conical singularities leaving
only black holes in the spectrum.

\section{Remarks}

We have exhibit in this note a set of solutions to the vacuum Einstein
equations which upon quantization yield the correct Bekenstein-Hawking
entropy.  In order to have a complete picture of the Hawking
evaporation process, we should consider also transitions between
states i.e., emissions of states. This type of
process will certainly need other correlation functions beyond the
partition function.  We shall not consider this problem here. However,
we would like to end by making a number of remarks which hopefully
will clarify the nature of the solutions that we have presented (at
least on the sphere), and may be the starting point towards a
dynamical modeling of the evaporation process in 2+1 dimensions.   

In particular, in section \ref{CS}, we shall argue that the
multivalued solutions --twisted sector-- arise in a natural way within
the Chern-Simons formulation of three-dimensional gravity. 

\subsection{The conformal algebra in the covering space}
\label{Covering}

Working with multivalued functions is often problematic and thus it is
convenient to pass to the covering space on which the stress tensor is
single valued. It is well-known in the theory of complex variables
that for any multivalued function of a complex coordinate
$z$, there exists a covering surface for which the given function is
single valued. In our case this covering is simply given by the
equation
\begin{equation}
z = u^q.
\label{zu}
\end{equation}
This transformation map the Riemann sphere (adS$_3$) into itself (up
to two singularities in the north and south poles where (\ref{zu}) has
fixed points).  For the torus, this transformation changes the genus
\cite{-Vafa}.  Since we do not have much control on the
three-dimensional solutions with higher genus 2d-boundaries, we shall
restrict ourselves in this section to the genus zero situation.  

We consider the metric (\ref{dsTT}) with $f$ given in
(\ref{fTq}) and $T$ in (\ref{Lnr}). Let us transform this metric with
(\ref{zu}). Two clarifications are necessary here. First, we shall
only
transform the complex coordinate $z$, not the radial coordinate. This
means that this transformation is not one of the residual
transformations mentioned before and we will obtain a different
metric. We could modify (\ref{zu}) including a radial dependence plus
adding the corresponding transformations for $\rho$ and obtain the
same metric defined on the covering space. For our purposes here this
is not necessary and would bring unnecessary complications. Second, we
shall treat $T$ classically and therefore we do not include any
Schwarztian derivative terms. Of course, if we did include the radial
reparametrization, there would be the classical Schwarztian derivative
piece.     

So, we simply plug (\ref{zu}) into (\ref{dsTT}) and find the new
metric 
\begin{equation}
 ds^2 = f du^2 +  \bar f d\bar u^2 + \left(q^2 |u|^{2q-2}\, e^{2\rho}
 + { f\bar f e^{-2\rho}\over q^2 |u|^{2q-2}} \right)du d\bar u +
 d\rho^2
\label{dsQQ}
\end{equation}
with 
\begin{equation}
 f(u) = {6q \over c}\, Q(u).
\label{hQ}
\end{equation}
A similar formula is valid for the anti-holomorphic side. The field
$Q$ is simply the conformally transformed version of $T$ under
(\ref{zu}),   
\begin{eqnarray}
Q(u) &=& \left({\partial z \over \partial u}\right)^2 T(u^q)
\nonumber\\
  &=:& \sum_{n\in Z} {Q_n \over u^{n+2}}
\label{Q}
\end{eqnarray}
and is single valued. The modes $Q_n$ are related to the modes of $T$
as \footnote{Here we have
used the monodromy decomposition: $$\sum_{n\in Z} {c_n\over x^{n/q}} =
\sum_{r=0}^{q-1} \sum_{m\in Z} {c_{mq+r} \over x^{m+r/q}}.$$}      
\begin{eqnarray} 
L_{n + r/q} = {1\over q}Q_{qn + r}.
\label{LQr}
\end{eqnarray}
These are exactly the generators introduced in Sec.
\ref{Quantization}
(see Eq. (\ref{LtQ})) which satisfy the Virasoro algebra (\ref{QQ})
with central charge $c/q$ . In this context they appear as generators
of the conformal symmetry acting on the covering space.    

The metric (\ref{dsQQ}) is an exact solution to the equations of
motion and it is single-valued.  However, it does not satisfy the
boundary conditions (\ref{aads}).  We shall see in Sec. \ref{CS} that
working in the Chern-Simons formulation one can easily deal with
metrics of the form (\ref{dsQQ}). 

How do we recover from (\ref{QQ}), the original Brown-Henneaux
algebra? The Virasoro algebra (\ref{QQ}) has the symmetry 
\begin{equation}
Q_n \rightarrow  e^{-2\pi i n/q} Q_n.
\label{S}
\end{equation}
It is direct to see that the subalgebra of the $Q's$ invariant
under (\ref{S}) is generated by the subset of generators
\begin{equation}
L_n = {1 \over q} Q_{qn} 
\label{LQ}
\end{equation}
and have central charge $c$. The sub-algebra spanned by the $L_n$'s
generates the single-valued conformal transformations acting on
$z=u^q$. Thus, they correspond to the Brown-Henneaux generators. 

This can also be seen as follows. We start with (\ref{Q}) and act on
this operator with the transformation (\ref{S}),
\begin{eqnarray}
Q(u) du^2 = \sum {Q_n  \over u^{n+2} } du^2 \longrightarrow  
\sum {e^{-2\pi i n/q} Q_n \over u^{n+2} }du^2  = Q(e^{2\pi i/q} u)
du^2 
\label{Qt}
\end{eqnarray}
Thus, (\ref{S}) act on the plane $u$ as  
\begin{equation}
u\rightarrow e^{2\pi i/q} u,
\label{tu}
\end{equation}
and the invariant coordinate is, as expected, $z=u^q$.  

\subsection{The action in the covering space}

We have considered twisted solutions to the equations of motion
associated to the action (\ref{I}), with the boundary condition
(\ref{aads}).  As shown in the last section, instead of working with a
multivalued function $T$, one can work with a single valued function
defined on a different space. In this section we study the action on
that space. This shed some light on the reduction in the central
charge in the covering space algebra. Due to conformal invariance, the
action in the covering space has the same local form as the action in
the original space. However, since the plane $u$ covers $q$ times the
original plane $z$, the coupling constant in the covering manifold is
$c/q$ instead of $c$.

This can be summarised in the following table. 
\begin{equation}
\begin{array}{c|c} 
  \mbox{Physical spacetime: } z  &  \mbox{Covering: } u=z^{1/q} 
    \\ \hline 
{\displaystyle {c \over 24\pi} \int_z (R+2) } & {\displaystyle  {c/q 
                             \over 24\pi }  \int_u (R+2) } \\ 
        \Downarrow      &     \Downarrow  \\
 \{L_{n+r/q},\ c \}& \{ Q_n,\ \  {\displaystyle {c  \over q}\}} 
             \end{array}
\label{table}
\end{equation}
It is useful to read this table backwards, from right to left. If we
restrict the action in the covering space  to the space of functions
invariant under (\ref{tu}) it reduces to an action of single valued
functions with coupling $c$ (recall that the sphere $u$ covers $q$
times the sphere $z$). This is our starting point and shows that the
coupling $c/q$ in the upper right corner is correct.   In the same
way, as showed in the last section, the generators $Q_n$ acting on the
invariant subspace reduce to the $L_n$'s ($r=0$).

This analysis posses the question of what is the coupling constant of
general relativity. In principle, the action in the plane $z$ is not
preferable in any form with respect to the action in the plane $u$. We
could then argue that $c/q$ should be related to Newton's instead of
$c$. The answer to this ambiguity, as in four dimensional gravity, is
found in the boundary conditions.  We define the action in the plane
$z$ with the boundary conditions (\ref{aads}) as the physical action. 
The transformation (\ref{zu}) changes the boundary conditions.  

\subsection{Chern-Simons formulation and the value of the coupling}   
\label{CS}

The Chern-Simons formulation provides a natural motivation to look for
multivalued solutions to the equations of motion.  Putting the
argument the other way around, within the Chern-Simons formulation,
the most natural solution to the equations of motion is (\ref{dsQQ}),
of which (\ref{dsTT}) is only a particular case.  Thus, when making
the conformal transformation (\ref{zu}) that brings (\ref{dsQQ}) to
the asymptotic form (\ref{aads}), one finds the multi-valued 
solutions introduced above.   

We consider here Euclidean three-dimensional gravity formulated as a
Chern-Simons theory \cite{Achucarro-,Witten88} for the group
$SL(2,C)$. The level of the Chern-Simons theory will be fixed below.  

The Chern-Simons action appropriated to the black hole problem is the
covariant one \cite{BBO,BM} supplemented by the chiral boundary
conditions, 
\begin{equation}
A^a_{\bar z}=0, \ \ \ \ \ \bar A^a_z=0.
\label{bcond}
\end{equation}
Recall that $A^a=w^a + ie^a$. At the boundary, the real and imaginary
parts of $A^a$ are linked by (\ref{bcond}). The non-zero components
$A_z$ and $\bar A_{\bar z}$ can be written as \cite{Baires},
\begin{equation}
A^a_z = 2w^a_z, \ \ \ \ \ A^a_{\bar z} = 2 w^a_{\bar z}.
\label{bcond1}
\end{equation}
Thus, (\ref{bcond}) reduces the $SL(2,C)$ gauge field at the boundary
to two {\it real} currents $A^a_z$ and $\bar A^a_{\bar z}$ satisfying
the affine $SU(2)$ algebra at level $k$.  

Conditions (\ref{bcond}) are not enough to ensure anti-de Sitter
asymptotics \cite{CHvD}.  Let $\{A^+,A^-,A^3\}$ and $\{\bar A^+,\bar
A^-,\bar A^3\}$ the components of the $SU(2)_k \times SU(2)_k$ affine
currents at the boundary (we omit here the subscripts $z$ and $\bar
z$). To ensure that the associated metric is asymptotically anti-de
Sitter, as in (\ref{aads}), the boundary conditions (\ref{bcond}) must
be supplemented with the reduction conditions \cite{CHvD} (see
\cite{B} for a different approach) 
\begin{eqnarray}
A^+=1,  \ \ & \ \ \bar A^-=1, \label{A+} \\
A^3=0, \ \  & \ \   \bar A^3=0 \label{A3}.
\end{eqnarray}
This leaves only $L:=A^-/k$ and $\bar L:=\bar A^+/k$ as arbitrary
functions.  It follows \cite{Polyakov90} that the affine $SU(2)_k$
algebra reduced by (\ref{A+}) and (\ref{A3}) yields for $L$ and $\bar
L$ the Virasoro algebra with central charge\footnote{In \cite{Baires},
the central charge was written as $c=-6k$. This sign can be changed by
a different orientation for the modes. Note that the Virasoro algebra
is invariant under $L_n \rightarrow -L_{-n}$  and $c\rightarrow -c$.
The following discussion is only a qualitative analysis and this sign
is irrelevant.} $c=6k$.  Identifying the level $k$ with the usual
value $k=l/4G$ yields $c=3l/2G$ and confirms that the generators
$L=A^-/k$ and $\bar L=\bar A^+/k$ are the Brown-Henneaux ones.  

As a motivation to a more general situation consider imposing only
(\ref{A3}).  The leading piece in the metric with these boundary
conditions is easily computed and yields,
\begin{equation}
ds^2  \sim e^{2\rho} A^+(u) \bar A^-(\bar u)  du d\bar u + d\rho^2.
\label{aadsA}
\end{equation}
It is clear that further restricting the gauge field with 
(\ref{A+}) leads to (\ref{aads}).  However, if we keep them as
arbitrary, but fixed, functions we discover that (\ref{aadsA}) is
precisely the asymptotic behaviour of (\ref{dsQQ}) with
\begin{equation}
A^+ = q u^{q-1}, \ \ \ \ \  \bar A^-= q \bar u^{q-1}
\label{Ag}
\end{equation}
Thus, the Chern-Simons formulation leads to (\ref{dsQQ}) in natural
way.   
 
It should be clear that, from the point of view of the Chern-Simons
formulation, there is no a priori reason to select the boundary
(\ref{aads}) instead of (\ref{aadsA}). The non trivial point is that
in the transition from (\ref{aadsA}) to (\ref{aads}) via $z=u^q$, we
have kept the twisted states, i.e., all branches of
$z^{1/q}$, by letting $T(z)$ to be a multivalued function. This is the
step that follows naturally in the Chern-Simons formulation because
there is no reason a priori to set $q=1$ in (\ref{Ag}). The
multivaluedness of $T(z)$ is then a simple consequence of the more
general boundary conditions in the covering space. 
 
The next question is what is the algebra leaving (\ref{aadsA})
invariant.  As a direct generalization of the above case, one can
prove that the restriction of the affine $SU(2)_k$ algebra via
(\ref{A3}) and (\ref{Ag}) lead for the combination $Q(u):=(1/k) A^+
A^- = (q/k) u^{q-1} A^-(u)$ the Virasoro algebra with a central charge
$c=6k$.  In this case, however, it is not correct to identify $Q$ with
the original Brown-Henneaux generators $L$, nor the central term $6k$
with the central charge (\ref{c}).  The generators $Q$ are the
operators introduced in Sec. (\ref{Covering}), acting on the
covering space with central charge $c/q$, and leave invariant the
generalized boundary conditions (\ref{aadsA}). We are then led to
identify the level $k$ as,  
\begin{equation}
6kq = c  = {3l \over 2G }
\label{k}
\end{equation}
in order to match the center of the algebra (\ref{QQ}).  Note also
that this is in agreement with (\ref{table}). The above Chern-Simons
theory, with the boundary conditions (\ref{aadsA}), is appropriated to
the action in the covering space with coupling $k = c/6q$. 

This issue brings in a problem of interpretation for the calculations
of partitions functions that have been done using the Chern-Simons
formulation, and its connection to chiral WZW
models\cite{Carlip97,BBO}.
The main assumption on this type of calculation is to quantize the
full gauge field $A^a$ without imposing and kind of reduction
conditions. However, as we have emphasized above, the values of $A^+$
and $\bar A^-$ change the global properties of the spacetime. The
coupling $k$ is not only related to Newton's constant but also to the
wrapping of the cover space into spacetime.      

There is another curious consequence of (\ref{k}). When quantising a
string theory on the adS$_3$ background, a formula can be given for
the spacetime conformal generators \cite{Giveon-}.  The associated
central charge is $c=6k_{st}p$ where $k_{st} = l^2/l_{st}^2$ is the
coupling of the $SL(2,\Re)$ WZW model, $p$ is the winding number
of the spacetime coordinate $X(\sigma)$ in the worldsheet, and
$l_{st}$ is the string length.  Newton's constant $G_3$ was associated
in \cite{Giveon-} to the winding number $p$ via
\begin{equation}
6 k_{st} p = {3l \over 2G_3}.  
\label{kst}
\end{equation}
The similarity between (\ref{k}) and (\ref{kst}) makes it tempting to
identify $k\sim k_{st}$ and $q \sim p$. This amounts to identify the
black hole boundary coordinate $z(u)$ with the string worldsheet
coordinate $X(\sigma)$ (compactified on a circle), i.e., treating the
covering coordinate as a string worldsheet. The fact that in the
quantum gravity approach one quantizes functions of $z$ and thus it
corresponds to string field theory has recently been discussed in
\cite{deBoer-}. Further, we note that in the sector with the maximum
degeneracy $c = q$ (see Sec. \ref{Quantization}), $k$ is of the order
of one which is also the value of $k_{st}$ is one expects the
Correspondence Principle \cite{Horowitz-} between black holes and
string states \cite{Mikovic} to work in three dimensions.   Finally,
the degrees of freedom of three-dimensional gravity, as formulated in
\cite{CHvD}, are described by a (constrained) non-chiral $SL(2,\Re)$
WZW model at level $k$. Following this line of reasoning, it would be
nice to see an explicit connection between the counting presented in
section \ref{Quantization}, and the string theory approach to black
holes.

\section*{Acknowledgments}

Useful discussions and correspondence with M. Asorey, F. Falceto, A.
Gomberoff, A. Mikovic, P. Navarro, J. Navarro-Salas and, especially,
A. Ritz are gratefully acknowledged. Financial support from CICYT
(Spain) grant AEN-97-1680, and the Spanish postdoctoral program of
Ministerio de Educaci\'on y Ciencia is also acknowledged.

\end{document}